\documentclass[aps,prl,superscriptaddress,twocolumn,floatfix,showpacs]{revtex4-2}
\usepackage[utf8]{inputenc}
\usepackage[T1]{fontenc}
\usepackage{graphicx,color}
\usepackage[utf8]{inputenc}
\usepackage{amsmath,,amssymb,bm,amsfonts,dsfont,mathrsfs,amsthm}
\usepackage{braket}
\usepackage{txfonts,comment}
\usepackage[colorlinks=true,linkcolor=blue,citecolor=blue,urlcolor=blue]{hyperref}
\usepackage{soul}
\usepackage{tikz}
\usetikzlibrary{shapes, arrows, positioning, calc, backgrounds}
\newcommand{\beginsupplement}
    {
    \makeatletter
    }

\begin{document}

\title{Modular Many-Body Quantum Sensors}

\author{Chiranjib Mukhopadhyay} 
\affiliation{Institute of Fundamental and Frontier Sciences, University of Electronic Sciences and Technology of China, Chengdu 611731, China}
\affiliation{Key Laboratory of Quantum Physics and Photonic Quantum Information, Ministry of Education,
University of Electronic Science and Technology of China, Chengdu 611731, China}
\author{Abolfazl Bayat}
\affiliation{Institute of Fundamental and Frontier Sciences, University of Electronic Sciences and Technology of China, Chengdu 611731, China}
\affiliation{Key Laboratory of Quantum Physics and Photonic Quantum Information, Ministry of Education,
University of Electronic Science and Technology of China, Chengdu 611731, China}
\begin{abstract}
Quantum many-body systems undergoing phase transitions have been proposed as probes enabling beyond-classical enhancement of sensing precision. However, this enhancement is usually limited to a very narrow region around the critical point. Here, we systematically develop a modular approach for introducing multiple phase transitions in a many-body system. This naturally allows us to enlarge the region of quantum-enhanced precision by encompassing the newly created phase boundaries. Our approach is general and can be applied to both symmetry-breaking and topological quantum sensors. In symmetry-breaking sensors, we show that the newly created critical points inherit the original universality class and a simple total magnetization measurement already suffices to locate them. In topological sensors, our modular construction creates multiple bands which leads to a rich phase diagram. In both cases, Heisenberg scaling for Hamiltonian parameter estimation is achieved at all the phase boundaries. This can be exploited to create a global sensor which significantly outperforms a uniform probe. 
\end{abstract}

\maketitle


\emph{Introduction.--} Quantum features allow us to build sensors which offer better precision than classical sensors employing the same amount of resources~\cite{helstrom1969quantum, holevo1984probabilistic, giovannetti2004quantum,boixo2008quantum, paris2009quantum,degen2017quantum,liu2019quantum,gu2008fidelity, gu2010fidelityreview, meyer2021fisher}. In many-body quantum probes, phase transitions are known to be a resource for achieving such enhancement in precision~\cite{venuti2007quantum,schwandt2009quantum,albuquerque2010quantum,gritsev2009universal, napolitano2011interaction, gu2008fidelity,greschner2013fidelity,frerot2018quantum,zhou2020quantum,rams2018limits,chu2021dynamic,mbeng2020quantum,di2022multiparameter,chu2023strong}. In fact, various manifestations of criticalities including second-order~\cite{zanardi2006ground,abasto2008fidelity,sun2010fisher,zanardi2008quantum,damski2013exact,rams2011scaling,salvatori2014quantum,yang2022super,fernandez2018heisenberg,montenegro2022sequential,ozaydin2015quantum, garbe2022critical,mirkhalaf2021criticality}, superradiant and Rabi type ~\cite{bin2019mass,dicandia2023critical,garbe2020critical, heugel2019quantum, lorenzo2017quantum,wu2021criticality,ying2022critical,tang2023enhancement,zhu2023rabi,garbe2022exponential}, topological~\cite{budich2020nonhermitian,koch2022quantum,free2022sarkar,zhang2023topological,srivastava2023topological}, dynamical~\cite{tsang2013quantum,macieszczak2016dynamical,carollo2018uhlmann}, Floquet~\cite{lang2015dynamical,mishra2021driving,mishra2022integrable}, continuous environmental monitoring~\cite{ilias2022criticality}, Stark localization ~\cite{he22023stark, Yousefjani2023long}, disorder-induced ~\cite{bhattacharyya2023disorderinduced,sahoo2023localization}, and boundary time crystals~\cite{montenegro2023boundary,cabot2023continuous} have already been exploited for sensing tasks. Experimental harnessing of criticality enhanced sensitivity has been achieved in NMR~\cite{liu2021experimental}, NV-centers in diamond~\cite{yu2019experimental,yu2022experimental}, trapped ions~\cite{ilias2023criticalityenhanced}, and Rydberg atoms~\cite{ding2022enhanced}. However, criticality enhanced sensitivity can be achieved only in a small region around the phase boundaries which makes it only useful for \emph{local sensing} tasks where the parameter of interest varies within a very narrow band. On the other hand, if the parameter varies over a wider range, known as \emph{global sensing}~\cite{montenegro2021global,mok2021optimal,rubio2021global}, the quantum advantage disappears quickly~\cite{montenegro2021global}, although adaptive protocols demanding real-time feedback control~\cite{salvia2023critical} have been proposed. Therefore, a central question is whether one can engineer many-body sensor probes successfully exploiting the broader phase diagram instead of only one (critical) point. 

In Refs.~\cite{derzhko2002quantum,derzhko2005regularly,tong2002quantum,delima2006the}, through conventional correlation functions and order parameters, it has been shown that periodically varying couplings and magnetic fields in $XY$ class of Hamiltonians may create multiple critical points. This leads to two natural questions - i) From a fundamental perspective, can one go beyond second order phase transitions and develop a generic recipe for creating multiple criticalities in diverse platforms?, and ii) From a quantum sensing perspective, do these additional critical points offer significant practical advantage in terms of harnessing the critical enhanced sensitivity over a wide range of parameters, thus opening the possibility for advancing global quantum sensor design? In this letter, we answer both questions in the affirmative by exploiting a systematic modular construction for multiplication of critical regions. In the symmetry-breaking case, multiple disordered islands emerge within the ordered phase. In the topological case, the modular sensor hosts multiple bands with different topological indices. Resulting additional critical regions all offer quantum-enhanced sensitivity with Heisenberg scaling. As an application, we show that these additional criticalities can be exploited towards a global quantum sensor which significantly outperforms same-sized uniform chains. 
\begin{figure}
    \centering
    \includegraphics[width=\linewidth]{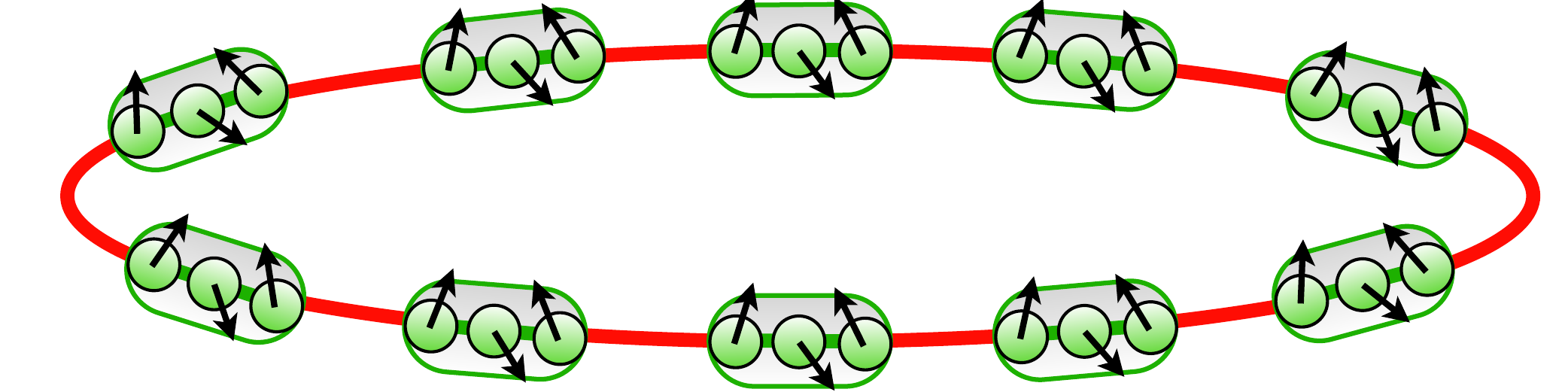}
    \caption{Modular quantum sensor design, where modules (grey with green borders) are connected via identical intercell couplings (red line) generally different from intracell couplings (green line).}
    \label{fig:local_xy_1}
\end{figure}
 
\emph{Quantum Estimation Theory.--} If an unknown parameter $\lambda$ is encoded in a quantum state $\rho_\lambda$, information about $\lambda$  can be obtained by measurement and subsequent post-processing of outcomes, via statistical estimation theory. The \emph{mean-squared error} of the estimator $\Lambda$ for parameter-value $\lambda$, is lower-bounded via the Cramer-Rao bound~\cite{nla.cat-vn81100, Rao1992, sidhu2020geometric, paris2009quantum} by $1/M Q$ after $M$-rounds, where $Q$ is the so-called \emph{quantum Fisher information (QFI)}. For a pure state $\rho_\lambda = |\psi_\lambda\rangle\langle\psi_\lambda|$, the expression for QFI simplifies as $Q = 4 \left( \langle \partial_\lambda \psi_\lambda | \partial_\lambda \psi_\lambda \rangle  - |\langle \psi_\lambda | \partial_\lambda \psi_\lambda \rangle|^2 \right)$~\cite{statistical1994braunstein}. We use this form of QFI for the ground state of many-body quantum sensors. The optimal measurement basis in general depends on the unknown parameter $\lambda$, which implies that the Cramer-Rao bound can only be achieved when $\lambda$ varies within a narrow range whose prior knowledge is provided, i.e., the \emph{local} estimation paradigm. However, in practice, the unknown parameter $\lambda$ may lie in a specified interval $[\lambda_0-\frac{\Delta \lambda}{2},\lambda_0 +\frac{\Delta \lambda}{2}]$, where $\Delta \lambda$ can be arbitrarily large, known as the \emph{global} estimation scenario. One can now define the \emph{global average uncertainty (GAU)} $G(\lambda_0|\Delta \lambda)$ as a figure of merit for quantifying the performance of the sensor~\cite{montenegro2021global,mok2021optimal} 
\begin{equation}
    G(\lambda_0|\Delta \lambda) = \int_{\lambda_0-\Delta \lambda/2}^{\lambda_0 +\Delta \lambda/2} \text{Var}(\Lambda') p(\lambda') d\lambda' \geq \int_{\lambda_0-\Delta \lambda/2}^{\lambda_0+\Delta \lambda/2} \frac{p(\lambda') d\lambda'}{Q(\lambda')},
    \label{eq:global_expression}
\end{equation}
where $p(\lambda)$ is the prior for the unknown parameter $\lambda$ henceforth assumed uniform to reflect maximum ignorance \footnote{We clarify that an apparently related, but subtly different to GAU, figure of merit due to Gill and Levit also exists in literature known as the Bayesian Cramer-Rao bound arising out of the well-known Van Trees inequality in mathematical statistics, however here we adopt a frequentist approach.}.
\\

\begin{figure}
    \centering
    \includegraphics[width=\linewidth]{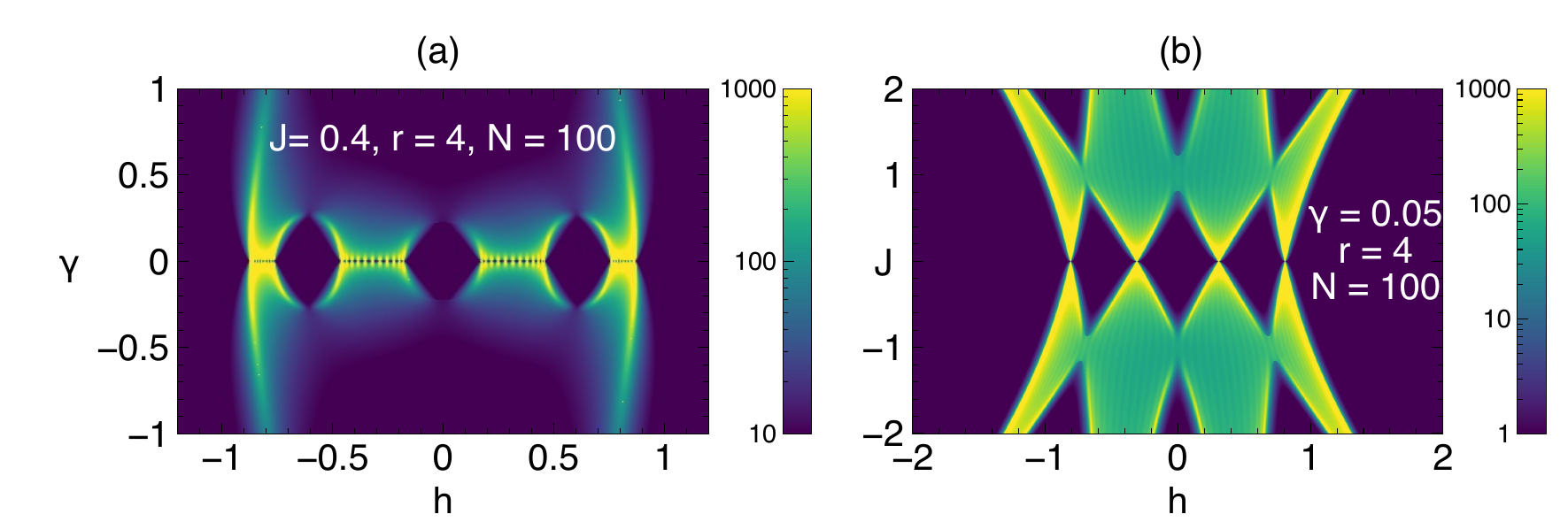}
    \caption{(a) Density plot of QFI for the modular sensor with period $r = 4$ and total system size $N = 100$ with anisotropy $\gamma$ and magnetic field $h$ at inter-cell coupling $J = 0.4$, (b) Density plot of QFI with inter-cell coupling $J$ and magnetic field $h$ for $\gamma = 0.05$ chain.}
    \label{fig:local_xy_1}
\end{figure}

\begin{figure}
    \centering
    \includegraphics[width=\linewidth]{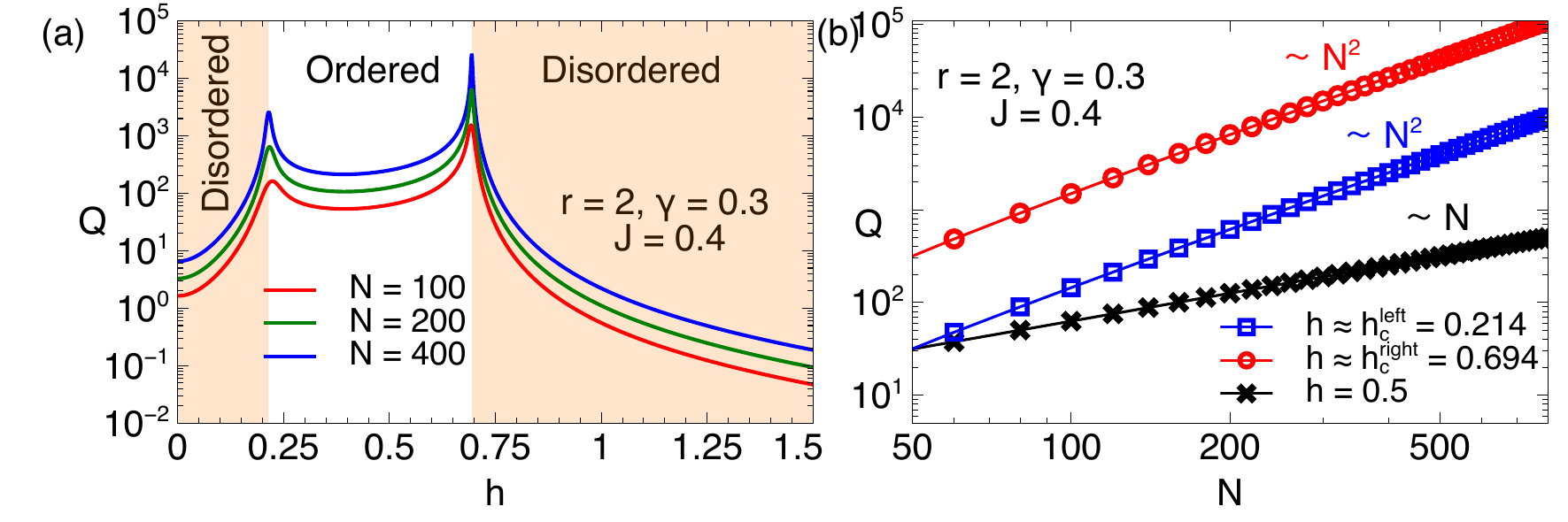}
    \caption{(a) QFI vs $h$ for dimerized sensors (i.e., period $r = 2$) for various system sizes $N$ and $J=0.4$. (b) Quadratic scaling of QFI with system size $N$ at the two phase boundaries (red when $h=0.214\approx h_{1}^{\text{left}}$ and blue when $h=0.694\approx h_{1}^{\text{left}}$ ), and linear scaling away from the phase boundaries (black) when $h=0.5$.}
    \label{fig:local_xy_2}
\end{figure}

\emph{Modular Probes as Local Sensors: I. Second Order Phase Transitions.--} We now introduce \emph{modular} many-body sensors, where couplings are uniform inside each module, but each module connects to other modules with a different inter-modular coupling strength. Specifically, we consider two concrete manifestations of modular many-body probes showing different types of phase transition, viz., second-order and topological. Let us consider the 1D anisotropic transverse-XY chain with  Hamiltonian $H = - \frac{1}{2} \sum_{i} J_{i} \left( (1+\gamma) \sigma_i^{x} \sigma_{i+1}^x + (1-\gamma) \sigma_i^{y}\sigma_{i+1}^{y} \right) + h \sum_{i} \sigma_i^{z}$, where the interaction $J_i$ is nearest-neighbor only, and the external magnetic field strength $h$ is to be estimated. In the modular construction {\color{black}(See Fig.~S1(a) of supplemental material for an illustration)}, the system is divided into several equal cells within which the coupling is uniform, viz., $J_{i}= J_{0}=1$, and neighboring cells are connected with tunable couplings $J_{i} = J$. Considering periodic boundary conditions, we assume the system of size $N = lr$ contains $l$-cells each with $r$ sites. This model can be solved by mapping to a free-fermionic Hamiltonian via a Jordan-Wigner mapping followed by Fourier and Bogoliubov transformations~\cite{delima2006the}. See SM, where we also provide a new and general expression for QFI in such free-fermionic systems. Critical points of this model at the thermodynamic limit can be found by a transfer matrix technique \citep{tong2002quantum,tong2006lee} as real roots of an $r$-th degree polynomial, resulting in up to $r-1$ paramagnetic islands in the ordered phase of the uniform system \footnote{see SM for the phase diagram illustration}. We now compute the QFI across the phase diagram with several paramagnetic islands. The QFI in the $(\gamma,h)$ phase plane for a fixed $J=0.4$ is shown in Fig.~\ref{fig:local_xy_1}(a). The QFI indeed peaks at each of the phase boundaries (which can be as large as $2r$ for low anisotropy $\gamma$) obtained theoretically, which suggests the possibility of quantum enhanced sensing near the phase boundaries. The same is obtained when we fix the anisotropy $\gamma = 0.05$ and consider the QFI in the $(J,h)$ phase plane in Fig.\ref{fig:local_xy_1}(b). To investigate the behavior of the QFI for different system sizes, in Fig.~\ref{fig:local_xy_2}(a), we take one slice of the phase diagrams above and plot the QFI for various system sizes as a function of $h$ for $J= 0.4$ and $\gamma = 0.3$ when the cell-size is $r=2$.  This choice of parameters results in $2r =4$ peaks \footnote{due to the symmetry of the phase diagram about $h=0$, we only illustrate QFI for half of the phase diagram with $h>0$} across the phase diagram which allows for a minimal representation of additional phase boundaries created by the modular sensor. As the figure indicates, increasing the system size $N$ results in larger QFI. To verify the criticality-enhanced sensitivity in Fig.~\ref{fig:local_xy_2}(b), we plot the QFI versus system size at and away from the phase boundaries. Indeed, the QFI shows Heisenberg scaling, i.e., $Q\sim N^2$, around \emph{all} the critical points, turning into standard scaling, i.e., $Q \sim N$, away from the criticality. Note that, the quantum enhanced scaling persists for \emph{all} the phase boundaries, creating multiple peaks across the phase diagram as module size~$r$ increase and thus significantly enhancing the tunability and flexibility of the probe to outperform classical sensors for a wider range of parameters.





\

\begin{figure} 
    \centering
    \includegraphics[width=\linewidth]{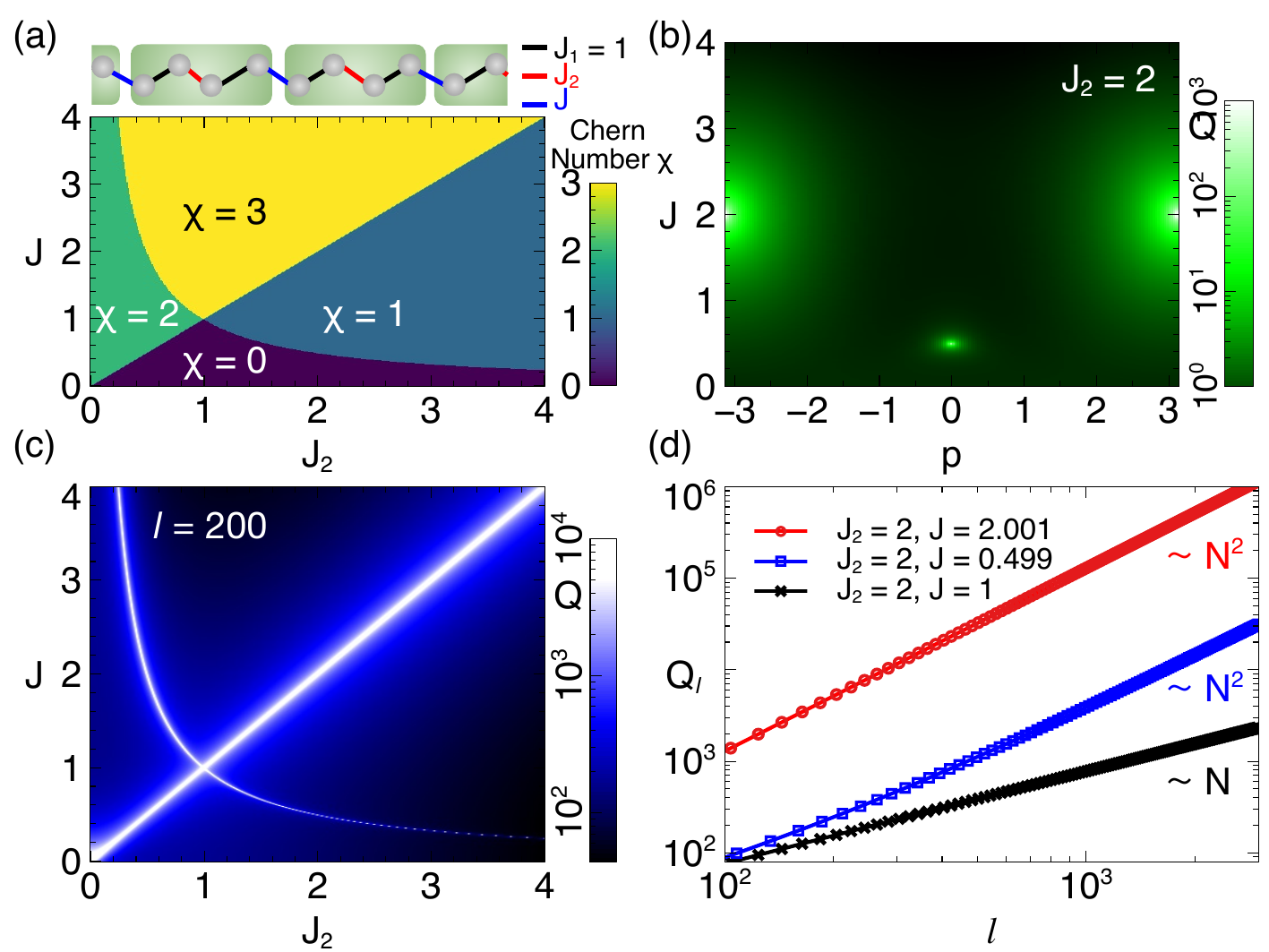}
    \caption{(a) Phase diagram of the simplest modular SSH sensor. (b) QFI for half-filled ground state at various points in the Brillouin zone,(c) Density plot of total QFI of the ground state at half-filling with couplings $(J_2,J)$ with $l=200$ modules, (d) Scaling of length-dependent component of total QFI at half-filling. }
    \label{fig:local_top}
\end{figure}

\emph{Modular Probes as Local Sensors: II. Topological Phase Transitions.--} We now show that the generation of multiple phase boundaries with modular sensors can be extended beyond second-order phase transitions by considering a modular topological quantum sensor based on the Su-Schrieffer-Heeger (SSH) model~\cite{su1979solitons}. In this case, each cell contains $r$ sites with alternating couplings $J_1$ and $J_2$. We put $J_1 =1$ as the unit of energy in our analysis and $J_2$ is the Hamiltonian parameter to be estimated. We consider the number of sites in each cell $r$ to be odd, as in this case, the modular sensor probe retains all the symmetries of the original SSH chain~\cite{lee2022winding}. The schematic of the modular probe is depicted in Fig.~\ref{fig:local_top}(a), in which different cells are connected with tunable inter-module couplings $J$. The corresponding Hamiltonian is $H = \sum_{i}  \left( J_1 c_{i,1}^{\dagger}c_{i,2} + J_2 c_{i,2}^{\dagger}c_{i,3} + ... + J c_{i,r}^{\dagger} c_{i+1,1} + h.c. \right)$,  where the system size is again $N = lr$. This Hamiltonian can be written as a direct sum of Bloch Hamiltonians $H_{\textrm{Bloch}}$, i.e., $H = \oplus_{p} H_{\textrm{Bloch}}(p)$, with $p$ being the crystal momentum, By diagonalizing the Bloch Hamiltonian, one finds that the energy bands occur in opposite signed pairs. Unlike the original SSH chain containing only two bands, the modular SSH probe contains $2r$-bands. As a result, by tuning $J$, one can close each of these $(2r-1)$ band gaps. For instance, for $J = J_2$ the gap between the middle two bands around zero-energy close at $p = \pm \pi/2$ of the Brillouin zone similar to the original SSH chain. Moreover, the other band gaps simultaneously close when $J = J_{2}^{-r+1}$  at the $p = 0$ of the Brillouin zone. While these two band gap closings suggest the presence of topological phase transitions, one may need to provide further evidence such as calculating the topological indices directly. Following the methodology of Ref.~\cite{lee2022winding}, for the smallest modular case $r=2$, we compute the winding number as a function of $J_2$ and $J$. The winding number is quantized and takes integer values between $0$ and $3$. The results are shown in Fig.~\ref{fig:local_top}(a), and the phase boundaries are indeed described by $J = J_{2}^{\pm 1}$.  
Let us now investigate the performance of the four-band modular SSH probe for this Hamiltonian parameter $J_2$ estimation problem. As Fig.~\ref{fig:local_top}(a) shows, the phase diagram is quite rich. Moreover, as predicted by the bulk-boundary hypothesis, non-trivial topological phases also host multiple edge-localized states~\cite{lee2022winding}. Since these edge states have the same exponentially decaying form as the generic edge-localized states \cite{alase2016exact,alase2017generalization,cobanera2018generalization}, one can repeat the analysis of Ref.~\cite{free2022sarkar} to show that the QFI of all of them scale quadratically with the total system size, i.e., $Q\sim N^2$. The case of many-body ground state at half-filling, i.e., all the lower bands occupied, is slightly more involved, as unlike the original SSH-chain \cite{free2022sarkar} we have to sum over more than one occupied band, see SM for more details. For the half-filling case,  Fig.~\ref{fig:local_top}(b) depicts the sum of the QFI for the two occupied lower bands in the ground state at $J_2 = 2$, in the plane of $(J,p)$.  As illustrated , the main contribution to the QFI comes from around $p = \lbrace\pm \pi, 0\rbrace $ points in the Brillouin zone, where the gap closes at the expected points  $J = J_2^{\pm 1}$ respectively. In order to get the total QFI of the ground state, one has to sum over all momenta and occupied bands. The results for the modular probe with $r=2$ and $N=400$ is shown in Fig.~\ref{fig:local_top}(c). Indeed, the peaks of the total QFI coincide with the topological phase boundaries found through the winding numbers shown in Fig.~\ref{fig:local_top}(a). In Fig.~\ref{fig:local_top}(d), we plot the ground state QFI at half-filling as a function of system size $N$ at various points of the phase diagram. Indeed, for two choices of parameters in the vicinities of two different phase boundaries, the QFI scales as~$\sim N^2$ asymptotically for both. In contrast, for a point far away from the phase boundaries, namely $(J,J_2) = (1,2)$, the QFI obeys shot noise scaling. This confirms that the modular sensor provides extra flexibility for tuning the probe to generate new topological quantum phase transitions with superior sensing capability.



\emph{Modular Many-body Probes Serving as Global Sensors.--} For a demonstration of the metrological power of newly created additional criticalities, we consider the paradigm of global sensing in which the unknown parameter $\lambda$ varies over an arbitrary wide range $\Delta \lambda$. In order to optimize the probe, we also assume that the unknown parameter $\lambda$ can be augmented with a known and tunable control value of the parameter $\lambda^{\textrm{ctr}}$ so that the total magnitude of the parameter to be sensed is $\lambda_{\textrm{tot}} = \lambda + \lambda^{\textrm{ctr}}$, where $\lambda \in [\lambda_{0} - \Delta \lambda/2, \lambda_0 + \Delta \lambda/2]$. As a figure of merit, we consider the GAU $G(\lambda_0 + \lambda^{\text{ctr}}|\Delta \lambda)$ defined as in Eq.~\eqref{eq:global_expression}. One can tune the probe by optimizing $\lambda^{\textrm{ctr}}$ to achieve minimum GAU as $ G^{\text{opt}}(\lambda|\Delta \lambda) = \min_{\lambda^{\textrm{ctr}}} G(\lambda_0 + \lambda^{\text{ctr}}|\Delta \lambda)$. Note that although the optimal control parameter $\lambda^{\text{ctr}^*}$ depends on $\lambda_0$, $\lambda_0 + \lambda^{\text{ctr}^*}$ is fixed for a given width $\Delta \lambda$.  \\

\begin{figure}
    \centering
    \includegraphics[width=\linewidth]{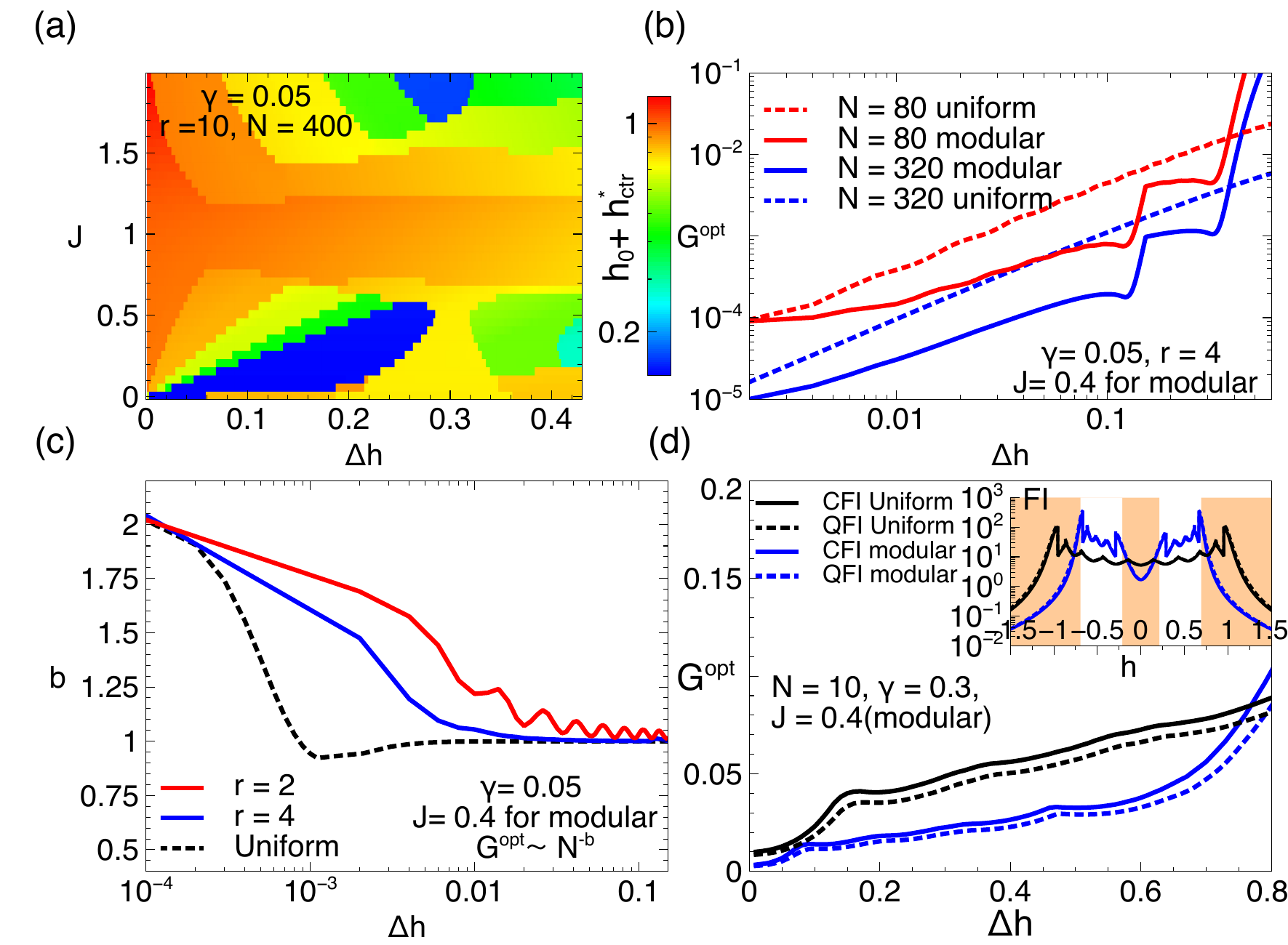}
    \caption{(a) $h_0 + h_{\text{ctr}}^{*}$ for optimized GAU $G^{\text{opt}}$. (b) GAU $G^{\text{opt}}$ vs. $\Delta h$ for various sizes $N$ of optimal modular sensors (solid, red for $N=80$, blue for $N=320$) when $J=0.4$  vs. optimal uniform sensors (dotted, red for $N=80$, blue for $N=320$) for $r=4$. (c) Decay of scaling exponent $b$ with width $\Delta h$, (d) Minimized GAU $G^{\text{opt}}$ vs width $\Delta h$ for uniform (black) and modular (blue) probes ($N=10$) for QFI-based ultimate bound $G$(dotted) and GAU $G^{\text{ach}}$ achieved through total transverse magnetization measurement (solid), inset : FI for local estimation vs $h$, orange(white) regions indicate (dis)ordered phases of modular system in thermodynamic limit.}
    \label{fig:global_xy}
\end{figure}

\emph{Modular Probes as Global Sensors: I. Second-Order Phase Transitions.--} We assume that the unknown magnetic field $h$ can be augmented with a known and tunable control magnetic field $h_{\textrm{ctr}}$ so that the total magnetic field $h_{\textrm{tot}} = h + h_{\textrm{ctr}}$, where $h \in [h_{0} - \Delta h/2, h_0 + \Delta h/2]$. Thus, in this case, our goal is optimizing $h_{\textrm{ctr}}$ to achieve minimum GAU as $G^{\text{opt}}(h|\Delta h) = \min_{h_{\textrm{ctr}}} G(h_0 + h_{\text{ctr}}|\Delta h)$. In Fig.~\ref{fig:global_xy}(a), we plot $h_0 + h_{\text{ctr}}^{*}$ in the $(J,\Delta h)$ plane.In the uniform chain, viz., $J=1$, the optimization simply shifts the probe to operate around its single critical point~\cite{montenegro2021global}, where $h_{0}+ h_{\text{ctr}}^{*}$ remains near unity. More interestingly, in the modular case, i.e., $J \neq 1$, the optimization becomes highly non-trivial and shifts the probe to operate around one of the several available critical regions as $\Delta h$ varies. Fig.~\ref{fig:global_xy}(a) confirms that the optimal probe configuration goes through abrupt changes as the width is increased. In order to quantitatively compare the performance of identically sized probes with different module sizes, in Fig.~\ref{fig:global_xy}(b) we investigate the effect of the total probe size over their performance through plotting the same quantity $G^{\text{opt}}(h_0 + h_{\text{ctr}}^{*}|\Delta h)$ as a function of $\Delta h$. While modular probes always outperform equally sized uniform probes, a modular probe of total size $N=80$ provides comparable performance with a uniform probe of $N=320$ sites for $\Delta h \approx 10^{-2}-10^{-3}$, reflecting significant size-efficiency for modular quantum sensors without any adaptive optimization strategies. This is also backed up by Fig.~\ref{fig:global_xy}(c), showing that for finite systems, the decay of scaling exponent from $\sim N^{-2}$ in the local estimation paradigm towards the shot-noise regime when regions far from criticality are encompassed, is more gradual for modular probes.

\emph{Achievable Global Precision.--} While the Cramer-Rao bound is theoretically tight for local estimation, the bound presented in Eq.~\eqref{eq:global_expression} is generally not tight, as the optimal measurement basis to saturate the Cramer-Rao bound may be different across the interval. Moreover, the optimal measurement may generally be highly complicated. Hence, by fixing a specific measurement setup $\Pi$, one can define an achievable global uncertainty for any parameter $\lambda \in [\lambda_0 - \Delta \lambda /2, \lambda_0 + \Delta \lambda/2] $ through this measurement setup as  $G^{\text{ach}} (\lambda_0| \Delta \lambda, \Pi) = \int_{\lambda_0 - \Delta \lambda/2}^{\lambda_0 + \Delta \lambda/2} p(\lambda^{'}) d\lambda^{'}/ Q_{C}^{\Pi} (\lambda^{'})$, where $Q_{C}^{\Pi}(\lambda)$ is the \emph{Classical Fisher Information (CFI)} for the measurement setup $\Pi$, i.e., $Q_{C}^{\Pi} (\lambda) = \sum_{i} (\partial_\lambda p_i)^2/p_i$, where the probabilities $p_i = |\langle i| \psi (\lambda)\rangle|^2$ over the measurement outcomes $\lbrace i \rbrace$. Clearly, this in general furnishes a higher average uncertainty compared to the bound in Eq.~\eqref{eq:global_expression}, even when optimised over the \textit{fixed} measurement settings. Now, we propose a simple fixed measurement of total transverse magnetization of the spin chain $Z = \sum_{j} \sigma_j^z$ with $(2N+1)$ outcomes for a probe of total size $N$, which has been previously shown to result in superlinear, albeit $\sim N^{1.5}$ \cite{salvia2023critical} scaling for uniform chains at the critical point. Remarkably, as we observe in the inset of Fig.~\ref{fig:global_xy}(d), for a small $(N=10)$ spin chain, this extremely simplified measurement is already very close to optimal for almost every value of the magnetic field $h$ away from criticality, and reproduces the same qualitative behaviour as the QFI discussed in the previous section. As confirmed in Fig.~\ref{fig:global_xy}(d), this translates to the global sensing capability as well. Thus, the modular probe, even when using a parameter-value independent simplified measurement setting, does indeed provide higher achievable precision, i.e., lower GAU, than the QFI-based bound on global precision obtainable through a uniform probe of the same size. \\

\begin{figure}
    \centering
    \includegraphics[width=\linewidth]{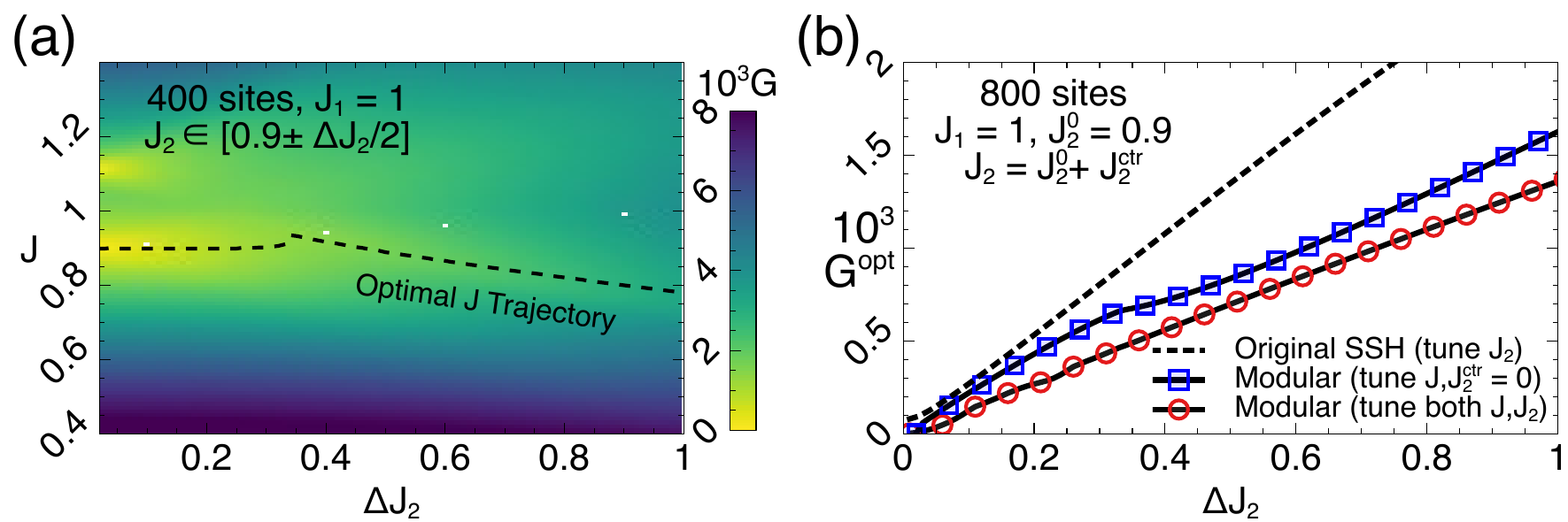}
    \caption{\textcolor{black}{(a) Heatmap of GAU $G$ for inter-modular coupling $J$ vs. width $\Delta J_2$ for $J_2 \in [0.9 \pm \frac{\Delta J_2}{2}]$ with $J_1 = 1, N = 400$. Black dashed line is optimal choice of $J$ to minimize $G$. (b) Minimum GAU $G^{opt}$ vs. $\Delta J_2$ for $N=800$ original SSH chain (black dashed), and $N=800$ modular SSH chains where further tuning $J_1$ is permissible (black solid, red circles) or inaccessible (black solid, blue squares).  }}
    \label{fig:global_ssh}
\end{figure}

\emph{Modular Probes as Global Sensors: II. Topological Phase Transitions.--} The goal is estimating the unknown coupling strength $J_2$ within the interval $J_2 \in [J_2^{0} - \frac{\Delta J_2}{2},J_2^{0} + \frac{\Delta J_2}{2} ]$ . We again assume a tunable augmentation of this coupling $J_2^{\text{ctr}}$, i.e., $J_2 \in [J_2^{0} + J_2^{\text{ctr}} - \frac{\Delta J_2}{2},J_2^{0} + J_2^{\text{ctr}} + \frac{\Delta J_2}{2} ]$. Moreover, to optimize the precision, we must also be able to tune the other couplings of the SSH chain. For the uniform SSH chain with couplings $J_1, J_2$ of the unit cell respectively, this entails tuning all $J_1$ couplings, i.e., half of the total couplings, collectively. For the modular SSH chain, two choices are present, the choice most similar to the uniform chain entails tuning both $J_1$ and $J$ couplings collectively. A sparser choice is to treat internal cell couplings $J_1 = 1$ as inaccessible and only tune the inter-modular coupling $J$. In Fig.~\ref{fig:global_ssh}(a), we depict the dependence of the GAU on the tunable coupling $J$ for fixed intra-cell coupling $J_1= 1$ and $J_2 \in [0.9 \pm \frac{\Delta J_2}{2}]$ with width $\Delta J_2$. For smaller widths, i.e., in the local sensing limit, the topological phase transition corresponding to $J \approx J_2^{0} = 0.9$ provides the minimal GAU, although the role of the other topological phase transition point at $J \approx 1/J_2^{0}$ in lowering the GAU is also evident. Remarkably, Fig.~\ref{fig:global_ssh}(b) confirms the minimum GAU attainable through the modular sensor to be far lower than the original SSH-chain based sensor of the same size \footnote{QFI of the two-band uniform SSH chain is obtained from the appendix of \cite{free2022sarkar}.}. This advantage only widens with the interval $\Delta J_2$, and holds true even when the intra-modular $J_1$ couplings are inaccessible.

{\emph{Generalization to Higher Dimensions.--} 
A higher-dimensional extension is straightforward since the key mechanism, inter-modular couplings effectively increasing the  unit-cell size, leading to more energy bands and thus more possibilities for gaps closing, remains unchanged. In the SM, we illustrate this with a concrete example of modular 2D SSH probes \cite{chen2018two}. Our results show that while the 2D uniform SSH model hosts only one critical point, the modular probe reveals several lines of criticality allowing significantly improved sensing capabilities across the phase diagram.}

\emph{Conclusion.--} We have proposed a modular approach for quantum many-body sensor design, which creates several new critical regions to be exploited for quantum-enhanced sensing, and leads to global quantum sensors requiring far fewer spins. In addition, modularity naturally increases the flexibility of the probe through adjustable intra-modular coupling strengths and module sizes. We note that the underlying transverse XY and SSH models have already been experimentally realized in different physical platforms, and with modularization via quantum charge-coupled devices \cite{pino2021demonstration} and matter-interconnect technologies \cite{akhtar2023high} becoming increasingly practical, our proposal is technologically feasible. Moreover, further improvement of modular probes via adaptive optimization and real-time feedback-control of couplings and measurement sequences remains available. 

\noindent \emph{Acknowledgements.--}   Authors acknowledge support from the National Key R\&D Program of China (Grant No.
2018YFA0306703), the National Natural Science Foundation of China (Grants No. 12050410253, No. 92065115, and No. 12274059), and the Ministry of Science and Technology of China (Grant No. QNJ2021167001L).

\bibliography{modular_prl_may}
\onecolumngrid
\beginsupplement
\section{Supplemental Material}

{ \color{black} \section{Introduction to Quantum Estimation Theory} 

Let us first briefly review quantum estimation theory (QET), as applied in this work. Historically, QET dates back to Helstrom \cite{helstrom1969quantum}, when he introduced the analog of the statistical Cramer-Rao bound in quantum theory, and Holevo's book \cite{holevo1984probabilistic} which interpretated the results in terms of Fisher information arising out of Born rule furnishing the probability model. Let us first review the Cramer-Rao bound, a key result in classical statistics, which states that the variance of the unbiased maximum likelihood estimator $\Lambda$ corresponding to the parameter value $\lambda$ after collecting $\mathcal{M}$-rounds of data is lower bounded as 

\begin{equation}
    \text{Var}[\Lambda] \geq \frac{1}{\mathcal{M} F_C (\lambda)}, 
\end{equation}

where the quantity $F_C$ is known as the Fisher Information and is given in terms of the underlying probability model, where the set of outcomes $\lbrace i\rbrace $ is associated with the corresponding probabiliity model $\lbrace p_i \rbrace$.  
\begin{equation}
    F_C = \sum_{i} p_i \left( \partial_\lambda \log p_i \right)^2
\end{equation}

In quantum theory, once the measurement bases are fixed, the probabilities are furnished by the Born rule, thus for a parameter $\lambda$ we want to estimate for a quantum state $\rho(\lambda)$, we maximise over Fisher Information obtained from the set of POVMs $\lbrace \Pi \rbrace$, i.e., 
\begin{equation}
    \mathcal{Q}(\lambda)  = \max_{\Pi} F_C (\Pi, \lambda)
\end{equation}

Alternatively, QFI can be expressed as expectation value of an operator, i.e., $Q = \text{Tr}[\rho_\lambda L_\lambda^2]$, where $L_\lambda$, the \emph{symmetric logarithmic derivative}, is defined in terms of the operator equation  $\partial_\lambda \rho_\lambda = \left( L_\lambda \rho_\lambda +  \rho_\lambda L_\lambda \right)/2$. Crucially, choosing the measurement basis as the eigenbasis of $L_\lambda$ saturates the Cramer-Rao bound, enabling us to use the QFI as the fundamental metric for ultimate metrological precision.  Braunstein and Caves' seminal paper \cite{statistical1994braunstein} expressed QFI in terms of statistical distance between quantum states, which for pure states $|\psi (\lambda)\rangle$ is simply given by $4 \left( \langle \partial_\lambda \psi (\lambda)|\partial_\lambda \psi (\lambda)\rangle - |\langle \psi(\lambda)|\partial_\lambda \psi (\lambda)\rangle|^2  \right) $ . This makes calculation of QFI far more easier than brute-force optimising of CFI over all measurement bases. Modern quantum sensing theory started using the language of QFI a decade latter, when Ramsey interferometry based optical sensing protocols were first discovered \cite{giovannetti2004quantum}. Many-Body sensing community, from its inception \cite{zanardi2008quantum, zanardi2006ground}, has always used (at least in the maximum-likelihood estimation setting, which is often implicit, unless a paper explicitly uses the Bayesian formalism, and even in this case, asymptotic precision is bounded by the QFI) the QFI as the key metric for precision. For a review on use of QFI for some many-body systems, we refer the interested reader to Ref.~\cite{gu2010fidelityreview}, while a more updated and comprehensive review will be available soon. A review on QFI in modern noisy intermediate-scale quantum devices may be found in Ref.~\cite{meyer2021fisher}. }\\

\section{QFI of the Modular Transverse XY Sensor.} Here, we provide a formula for the QFI of generic free-fermionic XY-like systems which has been used in this paper to calculate the QFI for the modular $XY$ sensors. For this let us take the generic $XY$-model again 
\begin{equation}
H = - \sum_{i,j} J_{i,i+1} \left( \frac{1+\gamma}{2} \sigma_i^{x} \sigma_{i+1}^x + \frac{1-\gamma}{2} \sigma_i^{y}\sigma_{i+1}^{y} \right) + \sum_{i} h_{i}\sigma_i^{z},
\end{equation} 
\noindent and use Jordan-Wigner transformation $\sigma_i^{x,y} = a_{i} \pm a_{i}^{\dagger}$ and $a_{i} = \exp\left[ -i\pi \sum_{j=1}^{i-1} c_{j}^{\dagger}c_{j} \right]c_{i}$ to map the system into a free-fermion problem with Hamiltonian
\begin{equation}
    H =\Psi^{\dagger} \tilde{H}~\Psi,
     \label{eq:Ham_supp}
\end{equation}
\noindent where $\Psi = \left[ c_1 c_2 ... c_1^{\dagger} c_{2}^{\dagger} ... \right]$ and $\tilde{H}$ is a $2N\times2N$ matrix with real entries expressed as~\cite{mbeng2020quantum}
\begin{eqnarray}
    \tilde{H} = \begin{bmatrix}
        A & B \\
        -B & -A
    \end{bmatrix}, \text{where}~A_{j,j} = h_{j}, A_{j,j+1} = A_{j+1,j} = -J_{j,j+1}/2,\nonumber \\ B_{j,j} = 0, B_{j,j+1} = -B_{j+1,j} = -\gamma J_{j,j+1}/2 \nonumber 
\end{eqnarray}

\noindent and boundary terms for periodic or anti-periodic boundary conditions are likewise defined in corners with the relevant signs. It can now be shown that the eigenvalues of $\tilde{H}$, and hence that of the original Hamiltonian, come in opposite signed pairs with eigenvectors $[U ~ ~ V]^{T}$ and $[V^{*} ~ ~ U^{*}]^{T}$ respectively. Now, we work out the QFI with respect to some parameter $\lambda$ for the $XY$-chain in this most general case. Let us denote $|\psi(\lambda)\rangle$ and $|\psi(\lambda+\epsilon)\rangle$ as ground states corresponding to parameter value being $\lambda$ and $\lambda+\epsilon$ respectively. The fidelity between these two ground states is given by Ohnishi's formula 
\begin{equation}
|\langle \psi(\lambda)|\psi(\lambda+\epsilon) \rangle| = \sqrt{\text{det} W},
\end{equation}

\noindent Where $W = U(\lambda)^{\dagger} U(\lambda+\epsilon) + V(\lambda)^{\dagger} V(\lambda+\epsilon)$. We can check that if $\epsilon = 0$, then from orthonormality of the eigenvectors, $W = \mathbb{I}$, and consequently the overlap is 1. The expression for QFI $Q(\lambda)$ now reads 

\begin{equation}
Q(\lambda) = \lim_{\epsilon \rightarrow 0} \frac{1-\sqrt{\text{det} W}}{\epsilon^2/2}
\end{equation} 

\noindent  We will use the shorthand notation $U(\lambda)=U, V(\lambda)=V$, $\text{det}W=\Delta$. Since $\Delta$ is very close to unity for small perturbations $\epsilon$, we can use perturbation expansion around $U,V$. Up to second order, this yields

\begin{eqnarray}
W   = \mathbb{I} + \epsilon \left( U^{\dagger}\frac{\partial U}{\partial \lambda} +  V^{\dagger}\frac{\partial V}{\partial \lambda} \right) + \epsilon^2 \frac{1}{2} \left( U^{\dagger}\frac{\partial^2 U}{\partial \lambda^2} +  V^{\dagger}\frac{\partial^2 V}{\partial \lambda^2} \right) \nonumber \\
= \mathbb{I} + \epsilon M_1 + \epsilon^2 M_2 
\end{eqnarray}

\noindent Using the matrix identity $\Delta = \exp\left(\text{Tr} \log W \right)$, and keeping terms up to second order, we have 

\begin{eqnarray}
\log \Delta & = \text{Tr} \left[ \log \left( \mathbb{I} + \epsilon M_1 + \epsilon^2 M_2   \right) \right] = \text{Tr} \left[\epsilon M_1 + \frac{\epsilon^2}{2} \left(M_2 - M_1^2  \right) \right].\nonumber 
\end{eqnarray}
\noindent It is easy to show that $\text{Tr} M_1 = 0$ follows from normalization and invariance of trace under adjoint. Hence, the overlap is expressed as 

\begin{equation}
\sqrt{\Delta} = e^{\frac{\epsilon^2}{2}\text{Tr}\left(M_2 - M_1^2\right)} \approx 1+ \frac{\epsilon^2}{2}\text{Tr}\left(M_2 - M_1^2\right) 
\end{equation}

 \noindent Which yields the following final expression of QFI of the parameter $\lambda$ in terms of eigenvectors of $\tilde{H}$ 
 \begin{eqnarray}
Q  = \text{Tr}\left[\left(U^{\dagger}\frac{\partial U}{\partial \lambda}\right)^2 - \frac{\partial^{2} U}{\partial \lambda^{2}} \right] + \text{Tr}\left[\left(V^{\dagger}\frac{\partial V}{\partial \lambda}\right)^2 - \frac{\partial^{2} V}{\partial \lambda^{2}} \right]  + 2 \text{Tr}\left[U^{\dagger}\frac{\partial U}{\partial \lambda} V^{\dagger} \frac{\partial V}{\partial \lambda}\right]
\end{eqnarray}

\noindent This form of QFI has, to the best of our knowledge, not been derived in literature. The decomposition of QFI into this form is evocative of interference effects between $U$ and $V$ set of eigenvectors, and should be explored in more detail. However, we can confirm that the presence of this interference term is \emph{not} solely responsible for the quantum enhancement in precision around criticalities. \\


\section{Phase Diagram of Modular Transverse XY sensor.}


Now, we briefly revisit the transfer matrix trick of determining the critical points of the modular $XY$ sensor following Ref.~\cite{tong2002quantum}. The eigenvalue equation in \eqref{eq:Ham_supp} can now be written in terms of the emergent excitations $\lbrace \eta \rbrace$ as 

\begin{eqnarray}
    \eta_{k}= \sum_{j=1}^{N} \left( U^{*}_{jk} c_{j} + V^{*}_{jk} c_{j}^{\dagger} \right)\\
    \eta_{k}^{\dagger} = \sum_{j=1}^{N} \left( V_{jk} c_{j} + U_{jk} c_{j}^{\dagger} \right),
\end{eqnarray}
\noindent such that  $ H = \sum_{k} \Lambda_k \left( \eta_{k}^{\dagger} \eta_k - \frac{1}{2}\right)$, where $\lbrace \Lambda_k, -\Lambda_k\rbrace$ are the energy eigenvalues. Now writing $\phi_{kj} = (U^{\dagger}+V^{\dagger})_{kj}/2$ and  $\xi_{kj} = (U^{\dagger}-V^{\dagger})_{kj}/2$, we can write the eigenvalue equation in the following recurrence form by expanding the eigenvalue equation through the Cayley-Hamilton theorem.

\begin{eqnarray}
\eta_k \phi_{kj} =   -J_{j-1,j} (1+\gamma) \phi_{k,j-1} - 2h_{j} \phi_{kj} - J_{j,j+1} (1-\gamma) \phi_{k,j+1} \\
\eta_{k} \xi_{kj} =  -J_{j-1,j} (1-\gamma) \xi_{k,j-1} - 2h_{j} \xi_{kj} - J_{j,j+1} (1+\gamma) \xi_{k,j+1}
\end{eqnarray}

\noindent At the critical point, the gaps close and the equations become decoupled. Solving for any one of them now means solving for the transfer matrix equation

\begin{equation}
    \Phi_{k,n} = \begin{bmatrix}
        \phi_{k,n+1} \\
        \phi_{k,n}
    \end{bmatrix} = \begin{bmatrix}
        -\frac{2h}{(1-\gamma)J_{n,n+1}} & -\frac{(1+\gamma)J_{n-1,n}}{(1-\gamma)J_{n,n+1}} \\
        1& 0
    \end{bmatrix}  \begin{bmatrix}
        \phi_{k,n} \\
        \phi_{k,n-1}
    \end{bmatrix} = M_{n} \Phi_{k,n-1}
\end{equation}

\noindent Non-trivial solution of this system is clearly possible iff  $\det(M \pm \mathbb{I}) = 0$, where $M =  M_{N}M_{N-1}.....M_{1} $. However, noting the periodicity of the $r$-periodic modular sensor with $l$ modules such that $N= lr$, this condition can be whittled down to $\det~(\tilde{M}^{l} \pm \mathbb{I}) = 0$ where $\tilde{M}$ is the product of transfer matrices within a single cell. By factorizing, the condition reduces to $\det~(\tilde{M} \pm \mathbb{I}) = 0$, where $\tilde{M}$ can now be written as 
\begin{equation}
    \tilde{M} = \begin{bmatrix} \frac{-2h}{J(1-\gamma)} & -\frac{1+\gamma}{J(1-\gamma)} \\ 1 & 0 \end{bmatrix} \begin{bmatrix} \frac{-2h}{1-\gamma} & -\frac{1+\gamma}{1-\gamma} \\ 1 & 0 \end{bmatrix}^{r-2} \begin{bmatrix} \frac{-2h}{1-\gamma} & -\frac{J(1+\gamma)}{1-\gamma} \\ 1 & 0 \end{bmatrix}
\end{equation}

\begin{figure}[t]
    \centering
    \includegraphics[width=\linewidth]{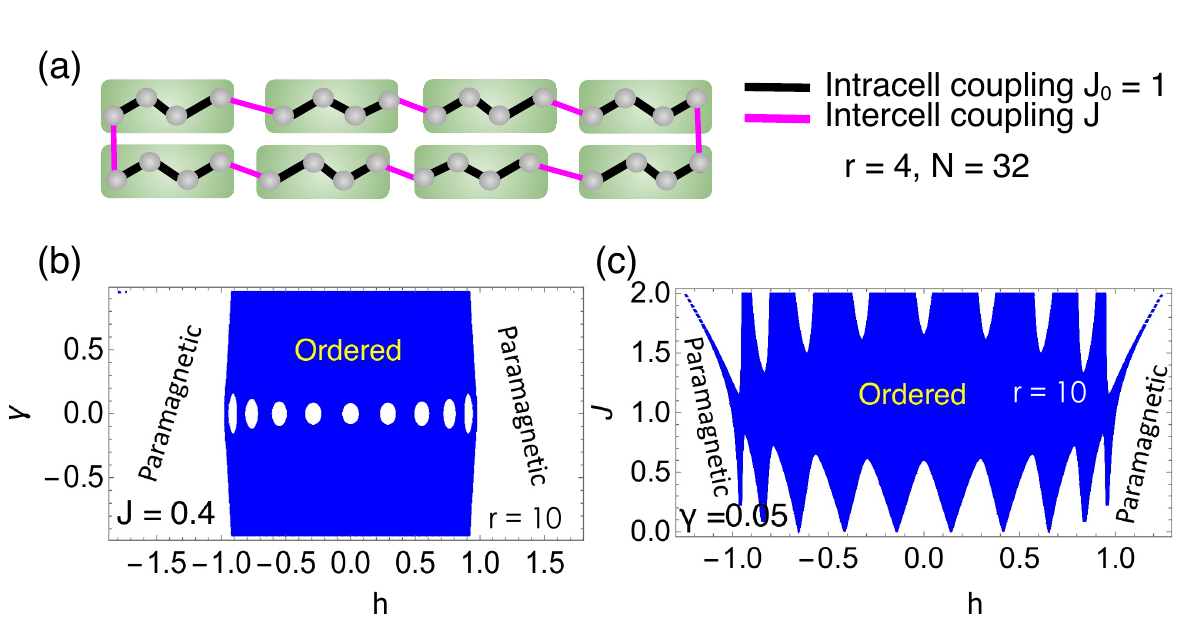}
    \caption{(a) Schematic of the modular construction of quantum sensor, (b) Phase Diagram of the modular sensor for $r=10$ at the thermodynamic limit in the $(\gamma,h)$ plane (left) and in the $(J,h)$ plane (right), notice the paramagnetic islands (white) inside the ferromagnetic (blue) phase. }
    \label{fig:supp_schematic}
\end{figure}

\noindent This condition is leads to a $r$-th degree polynomial equation, as alluded to in the main text. The resulting phase diagram in the $(h,\gamma)$ plane for a fixed $J = 0.4$ is depicted in Fig.~\ref{fig:supp_schematic}(b). It reveals that $r$ isolated paramagnetic islands emerge within the ordered phase of the corresponding uniform chain resulting in up to $2r$ phase boundaries. In the highly anisotropic regime, e.g., $\gamma = 1$ (transverse Ising case), the only effect is a slight shift in the original phase boundary at $h=1$ separating ordered and disordered phases. Here, we confine ourselves to studying the somewhat isotropic regime where $\gamma \rightarrow 0$. In this limit, we observe that $r$ paramagnetic islands emerge within the ordered phase. In fact, for any $|h|< 1$, it is possible to find a nearby critical point $h_c$ such that $|h-h_{c}| < \epsilon$, where $\epsilon \sim \mathcal{O}(1/r) $. Interestingly, as $h \rightarrow 1$, the paramagnetic islands become more dense such that $\epsilon \sim \mathcal{O}(1/r^2) $. In Fig.~\ref{fig:supp_schematic}(c), we depict the phase diagram in the $(J,h)$ plane for a fixed anisotropy $\gamma = 0.05$ where we again see the emergence of disordered regions within the ordered phase. As the figure shows, in order to realize multiple criticalities, intermodular couplings $J$ should be sufficiently different from the intracell coupling $J_0 =1$.\\ \\

\begin{figure}
    \centering
    \includegraphics[width=\linewidth]{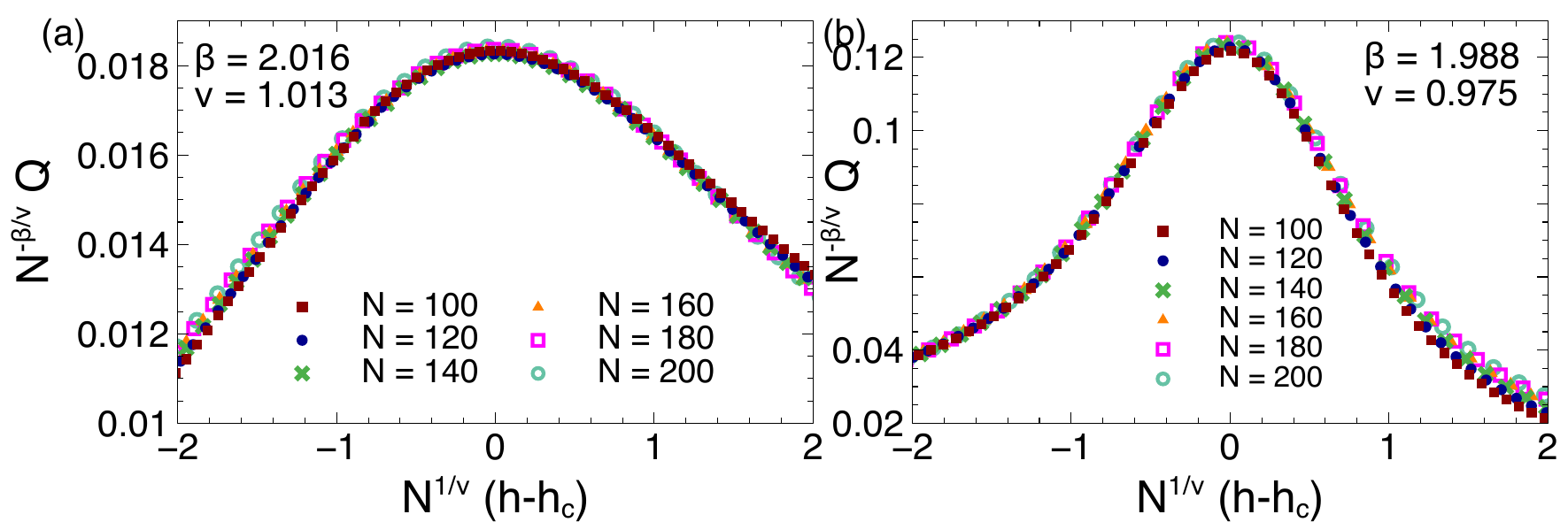}
    \caption{Finite size scaling data collapse for the two critical points depicted in the main text Fig.~\ref{fig:local_xy_2}: (a) $h_c = 0.214$ which gives $\beta = 2.016$ and $\nu = 1.013$; (b) $h_c = 0.694$ which gives $\beta = 1.988$ and $\nu = 0.975$.}
    \label{fig:supp_collapse}
\end{figure}

\noindent From a condensed matter perspective, one might be interested to determine the critical exponents associated with these extra critical points. In order to determine the critical exponents, we perform a finite size scaling analysis for the QFI across the criticalities observed in the figure above. The results are demonstrated in Fig.~\ref{fig:supp_collapse}(a)-(b) corresponding to the critical regions in Fig.~\ref{fig:local_xy_2} of the main text, when we consider an ansatz $Q = N^{\beta/\nu} f(N^{1/\nu}(h-h_c))$ for each of the critical points $h_c$. Note that the exponent $\beta$ determines the scaling at the critical point as $Q(h_c) \sim N^{\beta}$ and exponent $\nu$ quantifies the divergence of the length scale $\xi$ at the thermodynamic limit, viz., $\xi \sim |h-h_c|^{-\nu}$. By tuning $\beta$ and $\nu$, one can collapse the curves for $Q N^{-\beta/\nu}$ versus $N^{1/\nu}(h-h_c)$ on a single curve for various system sizes. Our analysis shows that for $h_c = 0.214$ ($h_c = 0.694$), one finds $\beta \approx 2.01 \pm 0.02$ ($\beta \approx 1.99 \pm 0.02$) and $\nu \approx 1.01 \pm 0.02$ ($\nu \approx 0.98 \pm 0.02$). The values obtained for the exponent $\beta$ through finite-size scaling analysis matches the results extracted directly from scaling analysis in Fig.~\ref{fig:local_xy_2}(b) of the main text.  Interestingly, the critical exponents from all these extra criticalities are roughly the same for the uniform XY-chain, \textcolor{black}{whose sensing performance was studied in Ref.~\cite{rams2011scaling}}. This shows that all these transitions fall within the same Ising-like universality class, and thus, the modular probe only increases the number of critical points without changing the nature of the transitions.

\section{QFI of Simplest Modular SSH-sensor at Half Filling.} Here, we sketch the calculation of QFI of the smallest 2-modular sensor at half filling. The Hamiltonian is given as 
\begin{equation}
    H = \sum_{i} \left( c_{i,1}^{\dagger} c_{i,2} + J_0 c_{i,2}^{\dagger}c_{i,3} + c_{i,3}^{\dagger}c_{i,4}+ J c_{i,4}^{\dagger}c_{i+1,1} + h.c. \right),
\end{equation}
\noindent and the corresponding Bloch Hamiltonian in momentum space is given by 
\begin{equation}
    H_{\text{Bloch}} = \begin{bmatrix}
        0 & 1 & 0 & J e^{-ip}\\
        1 & 0 & J_0 & 0\\
        0 & J_0 & 0 & 1 \\
        J e^{ip} & 0 & 1 & 0
    \end{bmatrix}
\end{equation}\\
Following the notation of Ref.~\cite{lee2022winding}, the eigenvalues $\Omega_i$ of the Bloch Hamiltonian come in opposite signed pairs $\pm \omega_1, \pm \omega_2 $, which are given by 
\begin{eqnarray}
\scriptsize \omega_1 = \sqrt{\frac{(J_0^2 + J^2+2) - \sqrt{(J_0^2 + J^2+2)^2 - 4 (1+J_{0}^2 J^2 - 2 J_{0}J \cos p)}}{2}} \\
\scriptsize \omega_2 = \sqrt{\frac{(J_0^2 + J^2+2) + \sqrt{(J_0^2 + J^2+2)^2 - 4 (1+J_{0}^2 J^2 - 2 J_{0}J \cos p)}}{2}} 
\end{eqnarray}
\noindent and the corresponding eigenvectors $|\psi_i\rangle$ for the $i$-th band is given by 
\begin{eqnarray}
    |\psi_i\rangle = \mathcal{N} \begin{bmatrix}
        \alpha_i \\
        \beta_i\\
        \gamma_i\\
        \delta_i
    \end{bmatrix}, \mathcal{N}= \frac{1}{\sqrt{|\alpha_i|^2 + |\beta_i|^2 + |\gamma_i|^2 + |\delta_i|^2}}, 
\end{eqnarray}
\noindent where 
\begin{eqnarray}
    \alpha_i = J (\Omega_i^2 - J_0^2)e^{-ip}+J_0 , \beta_i = -\Omega_i (J_0 + J e^{-ip}),\\
    \gamma_i = (\Omega_i^2 -1)+ JJ_0 e^{-ip}, \delta_i = \Omega_i (1+J_0^2 - \Omega_i^2)
\end{eqnarray}

Now, QFI with respect to the coupling strength $J_0$ for the $i$-th band is given by the fidelity susceptibility = $\langle \partial_{J_0} \psi_i | \partial_{J_0} \psi_i \rangle - |\langle \psi_i | \partial_{J_0} \psi_i \rangle|^2$. Plugging in the expression of the eigenvectors and simplifying, this leads to the following form of the QFI of the $i$-th band with crystal momentum $p$
\begin{eqnarray}
    Q_i (p) = \mathcal{N}^2 \left( |\partial_{J_0}\alpha_i|^2 + |\partial_{J_0}\beta_i|^2 + |\partial_{J_0}\gamma_i|^2 +|\partial_{J_0}\delta_i|^2 \right) \nonumber \\
    - \mathcal{N}^4 \left| \alpha_i^{*}\partial_{J_0}\alpha_i + \beta_i^{*}\partial_{J_0}\beta_i + \gamma_i^{*}\partial_{J_0}\gamma_i + \delta_i^{*}\partial_{J_0}\delta_i\right|^2
\end{eqnarray}

At the ground state configuration at half-filling, the lowest two bands (i.e. $i = 1,2$) corresponding to energies $-\omega_2, -\omega_1$ respectively, are fully occupied. Hence the total QFI of the system is given by integrating over the entire Brillouin zone and summing over these bands, i.e., 
\begin{equation}
    Q = \frac{1}{2\pi}\int_{-\pi}^{\pi} \left[ Q_1 (p) + Q_2(p)\right]dp
    \label{eq:1d_ssh_modular_qfi}
\end{equation}
\noindent In a finite lattice of $L$ modules, the integral over the Brillouin zone is replaced by a finite sum over crystal momenta $p = 2\pi k/L$ as $\frac{1}{2 \pi}\int \longrightarrow \sum_{k=0}^{L-1}$.  

{\color{black}\section{Dimensional Extension to 2D}

Let us now consider the generalization of the modular 1D SSH model in the main text to 2D. There are broadly two ways of doing so while preserving the inherent chiral symmetry of the SSH chain (see Ref. \cite{chen2018two} for detailed discussions). The first approach, the so-called type-I extension, is to glue together parallel SSH chains directed along the $x$-direction with staggered inter-chain hoppings. The second approach, the so-called type-II extension, is to shift the chain one step at a time for each layer to finally obtain  oblique edges. These two approaches are depicted for the modular SSH chain in Fig.~\ref{fig:supp_2d_extensions}. 

\begin{figure}
    \centering
    \includegraphics[width=\linewidth]{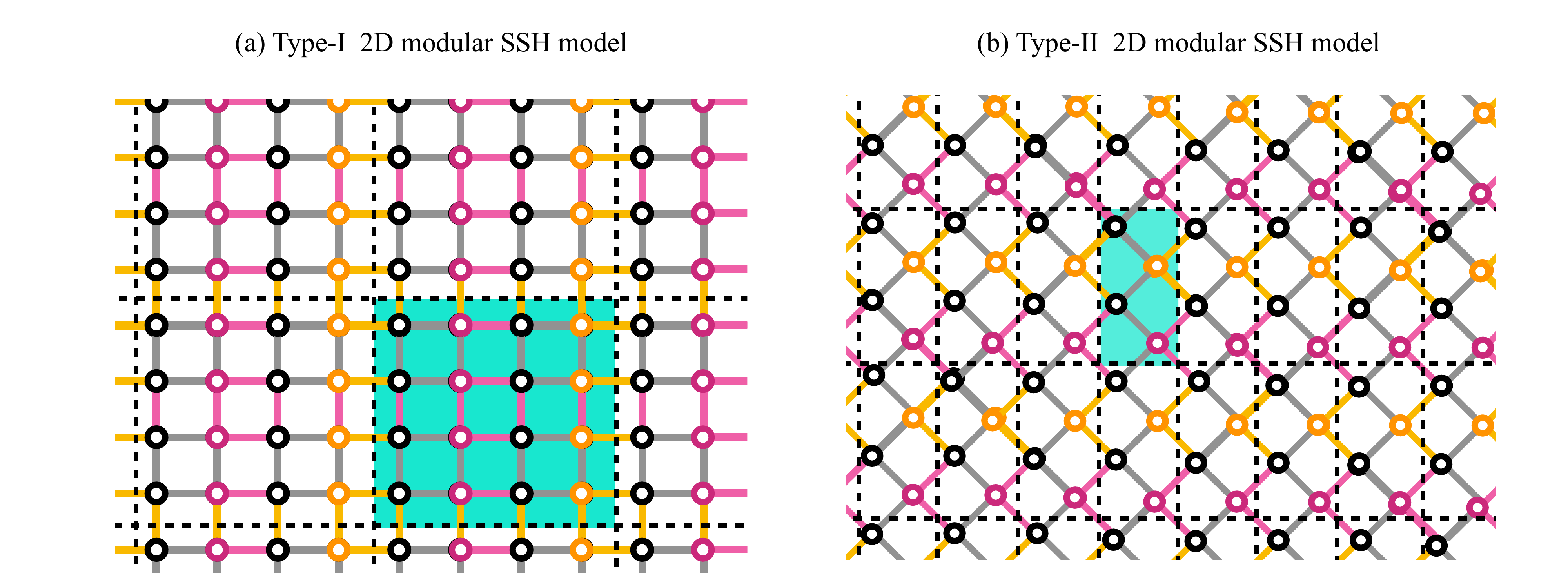}
    \caption{Dimensional extension of 1D modular SSH chain to 2D via (a) Type-I , and (b) Type-II extensions. Grey bonds indicate $J_1 = 1$, pink bonds $J_2$, golden inter-modular bonds $J$. Turquoise area indicates one unit cell.}
    \label{fig:supp_2d_extensions}
\end{figure}

Let us now work out the QFI analysis for the type-I extension. From Fig.~\ref{fig:supp_2d_extensions}(a), we note that there are sixteen sites per unit cell. From Ref.~\cite{chen2018two}, we know the Hamiltonian can be decomposed in terms of two 1D modular SSH chains along $x$ and $y$ directions, whose Hamiltonians are already given in the main-text, and which we denote as $H_{\text{SSH}}^{x}$ and $H_{\text{SSH}}^{y}$ respectively. That is the 2D SSH Hamiltonian is given by 

\begin{equation}
    H_{\text{SSH}}^{\text{2D}} = H_{\text{SSH}}^{x} \otimes \mathbb{I}^y + \mathbb{I}^x \otimes H_{\text{SSH}}^{y} 
\end{equation}

Given the additivity of the Hamiltonian, the QFI at many-body ground state of this Hamiltonian is simply

\begin{equation}
    |\psi^{\text{2D}}_{\text{ground}}\rangle = |\psi^{\text{1D}}_{\text{ground}}\rangle ^{\otimes 2},
\end{equation}
where $|\psi^{\text{1D}}_{\text{ground}}\rangle$ is the many-body ground state for the 1D modular SSH chain. Consequently the QFI is simply twice that of the 1D modular SSH chain analyzed before vide Eq.~\eqref{eq:1d_ssh_modular_qfi}. Since the QFI is shown to scale as $\sim N^2$ for an $N$-length 1D chain near the gap closing points, so does the QFI for the 2D SSH chain. 


Now let us concentrate on the type-II extension. In this case, the unit cell consists of four  sites, whereas for the non-modular SSH model, the unit cell consists solely of two sites \cite{chen2018two}. Thus, in real space, in terms of unit cell indices $\lbrace m\rbrace, \lbrace m'\rbrace$ along primitive vector directions, the Hamiltonian is given as
\begin{figure}
    \centering
    \includegraphics[width=\linewidth]{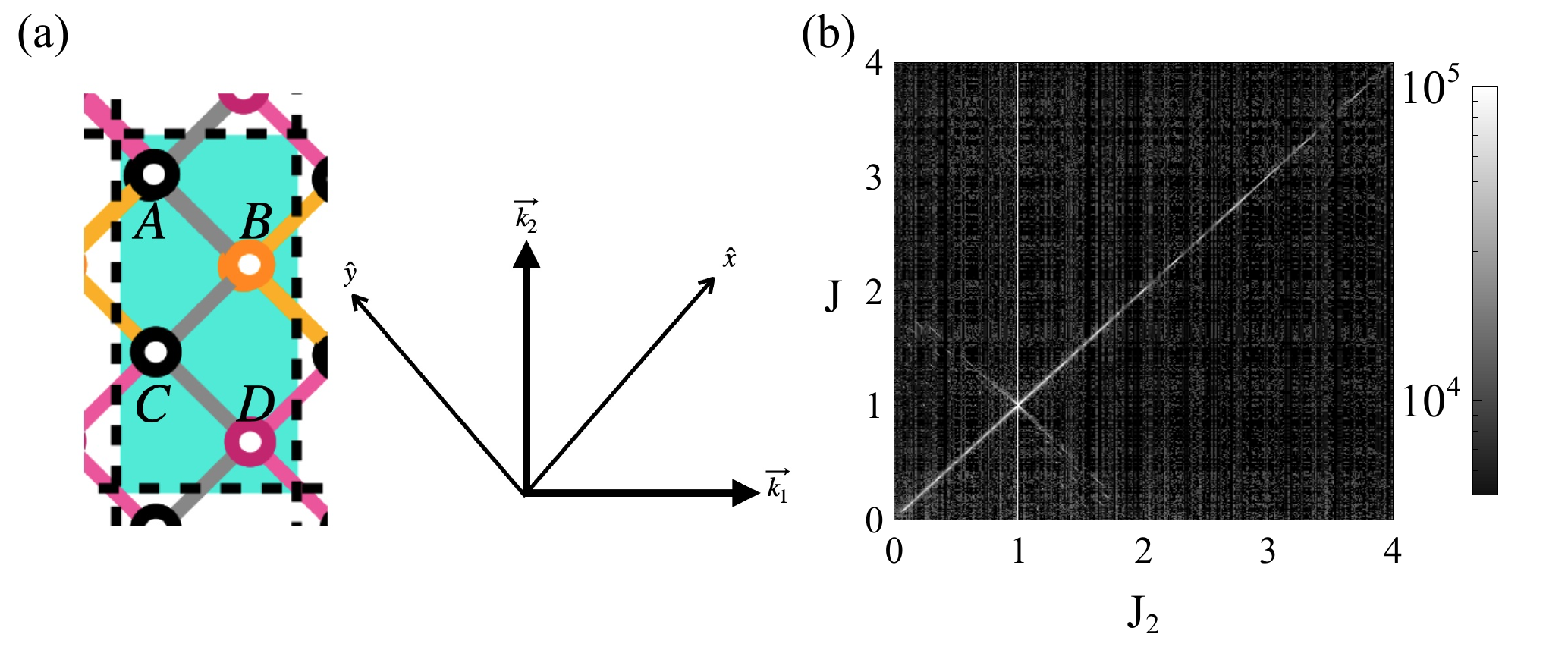}
    \caption{(a) One unit cell magnified for the type-II extension with site indices labelled. (b) Heatmap of QFI wrt coupling $J_2$ in $(J,J_2)$ space when $J_1=1$, for a $60\times60$ 2d modular SSH lattice built with the type-II extension showing QFI maximized along $J = J_2$, $J_2 = 1$, and $J + J_2 =2$ directions. In contrast, $J = J_2 = J_1 = 1$ is the sole critical point of the corresponding uniform 2D SSH chain. }
    \label{fig:supp_2d_ssh_qfi}
\end{figure}
\begin{eqnarray}
H = \sum_{m_i}\sum_{m'_j} J_1 |m_i,m'_j; A\rangle\langle m_i,m'_j; B| + J_1 |m_i,m'_j; B\rangle\langle m_i,m'_j; C| + J_1 |m_i,m'_j; C\rangle\langle m_i,m'_j; D| \nonumber\\ 
+J_1 |m_i,m'_j; A\rangle\langle m_i,m'_j+1; D| + J_2 |m_i,m'_j; A\rangle\langle m_i+1,m'_j+1; D| + J |m_i,m'_j; B\rangle\langle m_i+1,m'_j; A| \nonumber \\
 + J |m_i,m'_j; B\rangle\langle m_i+1,m'_j; C| +  J_2 |m_i,m'_j; D\rangle\langle m_i+1,m'_j; C| + h.c.
\end{eqnarray}
Consequently, the $4 \times 4$ Bloch Hamiltonian in the momentum space is given by 

\begin{equation}
    H_{\text{Bloch}} = \langle k_1, k_2 |H| k_1, k_2\rangle = \begin{bmatrix}
    0 &J_1+ J e^{-ik_1} &0 & J_1 e^{-ik_2} + J_2 e^{-ik_1-ik_2} \\
    J_1+ J e^{ik_1} &0 &J_1 + J e^{-ik_1} &0 \\
    0 &J_1+ J e^{ik_1} &0 &J_1 + J_2 e^{-ik_1} \\
    J_1 e^{ik_2} + J_2 e^{ik_1+ik_2}  &0 &J_1 + J_2 e^{ik_1} &0 \\
    \end{bmatrix}
    \label{eq:2d_bloch_ham}
\end{equation}
where the unit-vectors $k_1 = \frac{2\pi m}{N_1}$ and $k_2 = \frac{2\pi m'}{N_2}$ ; $m\in \mathbb{Z} \cap [0,N_1 -1] ; m'\in \mathbb{Z} \cap [0,N_2-1] $ along orthogonal unit cell primitive vector directions, assuming the lattice stretches for $N_1$ and $N_2$ unit cells along these two directions. One can now express the ground state QFI as sum over lower two bands, i.e., 

\begin{equation}
    \mathcal{Q} = \sum_{m = 0}^{N_1 -1}\sum_{m' = 0}^{N_2 -1}   \mathcal{Q}_1 (m,m') + \mathcal{Q}_2 (m,m'),
\end{equation}

By setting $J_1 = 1$ as before and diagonalizing the above Hamiltonian, we see that the energy gap closings occur for $J_2 = 1, J_2 = J$, and $J_2 + J = 2$. This is in sharp contrast to the non-modular 2D SSH model ($J = J_2)$, where the gap only closes where hopping terms become equal ($J_1 = J_2 =1$) corresponding to a single point, i.e., $(1,1)$, in the $J-J_2$ phase diagram. However, this is similar to the behaviour for the 1D modular SSH model, where the gap instead closes for $J_2 = J^{\pm 1}$ in the main text. Moreover, as we observe from Fig.~\ref{fig:supp_2d_ssh_qfi}, the QFI is indeed maximized along these directions. Thus, the core thesis of this work, improved quantum sensing capability with modular probes by opening up new regions of the phase diagram for gap-closing with tunable couplings, is again validated in 2D.

}
\end{document}